\documentclass[amssymb,twocolumn,pra,reprint,amsmath,amssymb,aps,floatfix,superscriptaddress,nofootinbib]{revtex4-2}

\usepackage{graphicx}
\usepackage{dcolumn}
\usepackage{bm}
\usepackage{epstopdf}
 \usepackage[dvipsnames]{pstricks}
 \usepackage{epsfig}
  \usepackage[normalem]{ulem} 
 \usepackage{pst-grad} 
 \usepackage{pst-plot} 
 \usepackage[space]{grffile} 
 \usepackage{etoolbox} 
\usepackage{booktabs}
\usepackage{boxhandler}
\usepackage{amsmath}
\usepackage{bbold}
\usepackage{natbib}
\usepackage[whole]{bxcjkjatype}
\usepackage{tikz}
\usepackage{soul}
\usepackage{mathtools}
\usetikzlibrary{arrows.meta}
\usepackage{hyperref}
\hypersetup{
  colorlinks   = true,    
  urlcolor     = blue,    
  linkcolor    = blue,    
  citecolor    = blue      
}

\usepackage[caption=false]{subfig}
\usepackage{floatrow}
\usepackage[utf8]{inputenc}
\usepackage{braket}
\usepackage[cmintegrals]{newtxmath}
\usepackage{multirow}
\usepackage{enumitem}
\usepackage{pifont}

\newcommand{\Rub}[1]{\textcolor{Blue}{{#1}}}

\begin{document}

\title{Sub-GHz Breathing Dynamics of Magnetic Hopfions}

\author{Felipe Tejo}
\affiliation{Universidad Central de Chile, Vicerrectoría de Investigación, Innovación y Postgrado, Santiago, Chile.}
\author{Rubén M. Otxoa}
\email{ro274@cam.ac.uk}
\affiliation{Hitachi Cambridge Laboratory, J. J. Thomson Avenue, Cambridge CB3 0HE, United Kingdom.}
\begin{abstract}
Magnetic hopfions are three-dimensional topological solitons whose static stability has recently been confirmed in experiments, yet their dynamical modes remain largely unexplored. Here we combine micromagnetic simulations and analytical modelling to characterise the fundamental breathing excitation of hopfions. We show that the breathing mode corresponds to a coherent oscillation of both the hopfion core diameter and the shell width, while preserving the topological charge. An analytical domain-wall interaction model explains the weak field dependence of the shell thickness and yields a closed-form expression for the restoring stiffness. From this curvature and the collective-coordinate inertia, we derive an estimated breathing frequency in excellent agreement with micromagnetic spectra. The ability to capture the hopfion dynamics quantitatively from material constants highlights a direct route to experimental detection by ferromagnetic resonance or Brillouin light scattering, and establishes a framework for frequency-encoded control in reconfigurable spintronic devices.

\end{abstract}
\flushbottom
\maketitle
%
%
\thispagestyle{empty}

\section*{Introduction}
Topological solitons are stable and localized field configurations characterised by integer-valued invariants that remain unchanged under continuous deformations \cite{manton2004chapter4}. Within this broad family, hopfions constitute a unique three-dimensional realisation, distinguished by a nonzero Hopf index that quantifies the linking of magnetization preimages in real space~\cite{rybakov2022magnetic,kobayashi2014torus}. Originally proposed in field theories \cite{korepin1975quantization,faddeev1997stable}, hopfions have since been investigated across diverse physical systems, including fluid dynamics~\cite{kleckner2013creation,kleckner2016superfluid}, nonlinear optics~\cite{dennis2010isolated}, chiral liquid crystals~\cite{ackerman2015self, ackerman2017static}, cosmology~\cite{thompson2015classification}, and Bose--Einstein condensates~\cite{kawaguchi2008knots,hall2016tying,annala2024optical}. These studies highlight their fundamental and interdisciplinary relevance. In nanomagnetism, the existence of hopfions has only recently been confirmed experimentally \cite{kent2021creation,zheng2023hopfion}; however, their internal dynamics remain largely unexplored. The topological nature of hopfions arises from the nontrivial mapping \(\mathbb{R}^3 \cup \{\infty\} \to \mathbb{S}^2\), which differs fundamentally from the skyrmion number used to describe two-dimensional (2D) magnetic textures \cite{manton2004topological,borisov2011three,sutcliffe2018hopfions}. The Hopf-index is formally defined as
\[
H = \frac{1}{(4\pi)^2} \int_V \mathbf{F} \cdot \mathbf{A} \, dV,
\]
where \(F_i = \tfrac{1}{2}\,\varepsilon_{ijk}\,\mathbf{m}\cdot(\partial_j \mathbf{m}\times \partial_k \mathbf{m})\), with \(i,j,k \in \{x,y,z\}\) and \(\varepsilon_{ijk}\) the Levi--Civita tensor. In this definition, \(\mathbf{F}\) is constructed from the magnetization texture through derivatives of \(\mathbf{m}\), while \(\mathbf{A}\) denotes a vector potential such that \(\mathbf{F} = \nabla \times \mathbf{A}\). Physically, \(\mathbf{F}\) can be interpreted as a gyrovector density~\cite{thiele1973steady}, an emergent magnetic field~\cite{zhang2009generalization}, or a local topological charge density~\cite{braun2012topological}. 
This richer invariant suggests the existence of dynamical behaviours that transcend those encountered in two-dimensional textures, since hopfions require full three-dimensional confinement and a more intricate competition of energies, in contrast to two-dimensional solitons such as skyrmions and vortices. In two-dimensional systems, low-frequency eigenmodes—including breathing and gyration—are well established both in skyrmions \cite{kim2014breathing, garcia2016skyrmion} and in other magnetic textures such as vortices \cite{ding2014higher, petit2012commensurability,  otxoa2011nanocontact,sanches2014current, kamionka2011influence} and domain walls (DWs) \cite{borys2016spin, albert2016domain, fattouhi2025dissipative,bolte2006spin,tatara2020collective}. These collective excitations form the foundation for proposals in microwave devices, neuromorphic computing, and magnonic logic\cite{khitun2025magnonic, ababei2021neuromorphic,jiang2019physical,velez2025chaotic,nakane2021spin,otxoa2023tailoring,lee2022reservoir,pinna2020reservoir,sun2023experimental}. By contrast, in hopfions no equivalent body of work yet exists, even though their three-dimensional topology naturally points to the possibility of novel, decoupled dynamical degrees of freedom that are inaccessible to two-dimensional textures. In particular, whether hopfions can host rigid-core oscillations—where the topological core remains intact while the surrounding structure deforms coherently—has yet to be addressed.
In this context, the response of hopfions to external magnetic fields \cite{liu2018binding, sutcliffe2017skyrmion, tai2018static}, spin currents \cite{wang2019current, liu2022emergent, liu2020three, liu2020erratum, gobel2020topological}, or thermal gradients raises fundamental questions regarding their robustness and potential for device implementation. Several studies have examined the resonance spectra of confined hopfions subjected to magnetic pulse excitations \cite{khodzhaev2022hopfion, raftrey2021field}, focusing on spectral features and their evolution across phase transitions, such as those between hopfions, torons, and other metastable states \cite{bo2021spin, li2022mutual}. Although these investigations reported a variety of excitation modes, the dynamical landscape of hopfions remains far from complete. 
Here, by combining high-resolution micromagnetic simulations with a tractable analytical model, we demonstrate that a confined magnetic hopfion supports a field-tunable rigid-core breathing mode.
In this work, we report that a magnetic hopfion confined within a rectangular ferromagnetic nanodot exhibits a diameter that can be continuously and reversibly tuned by an out-of-plane magnetic field, while its internal core width remains remarkably constant. This geometric decoupling between external deformation and internal rigidity (absent in 2D skyrmions) constitutes a new form of \textit{elasticity} in a magnetic soliton. By exploiting this separation, we identify through time-resolved micromagnetic simulations and frequency domain analysis a robust sub-GHz breathing mode localized at the boundary of the hopfion, in which the hopfion undergoes coherent radial oscillations without altering its internal topological structure. Moreover, we show that the hopfion internal core width responds asymmetrically to the magnetic field polarity, as a result of the competition between the DW tensions forming the hopfion, the exchange-mediated repulsion among them, and the Zeeman energy compression of the internal DWs that constitute the internal structure of the hopfion. This interplay not only governs the texture's internal deformation, but also enables a low-frequency, spatially selective breathing mode. In addition, our analytical domain-wall interaction model reproduces the same low-frequency breathing excitation, providing a closed-form estimate of the restoring stiffness and predicting a breathing frequency, $f_0$ in quantitative agreement with the micromagnetic spectra. We envision that this new excitation mode should be directly detectable in experiments using ferromagnetic resonance \cite{van2023x} or Brillouin light scattering techniques \cite{swyt2024brillouin}.
These findings unveil a new degree of tunability and dynamic functionality in magnetic hopfions, positioning them as promising candidates for reconfigurable information carriers in three-dimensional spintronic devices.
\section*{Micromagnetic simulations}
\begin{figure*}[!ht]
    \centering
    \includegraphics[width=1.0\linewidth]{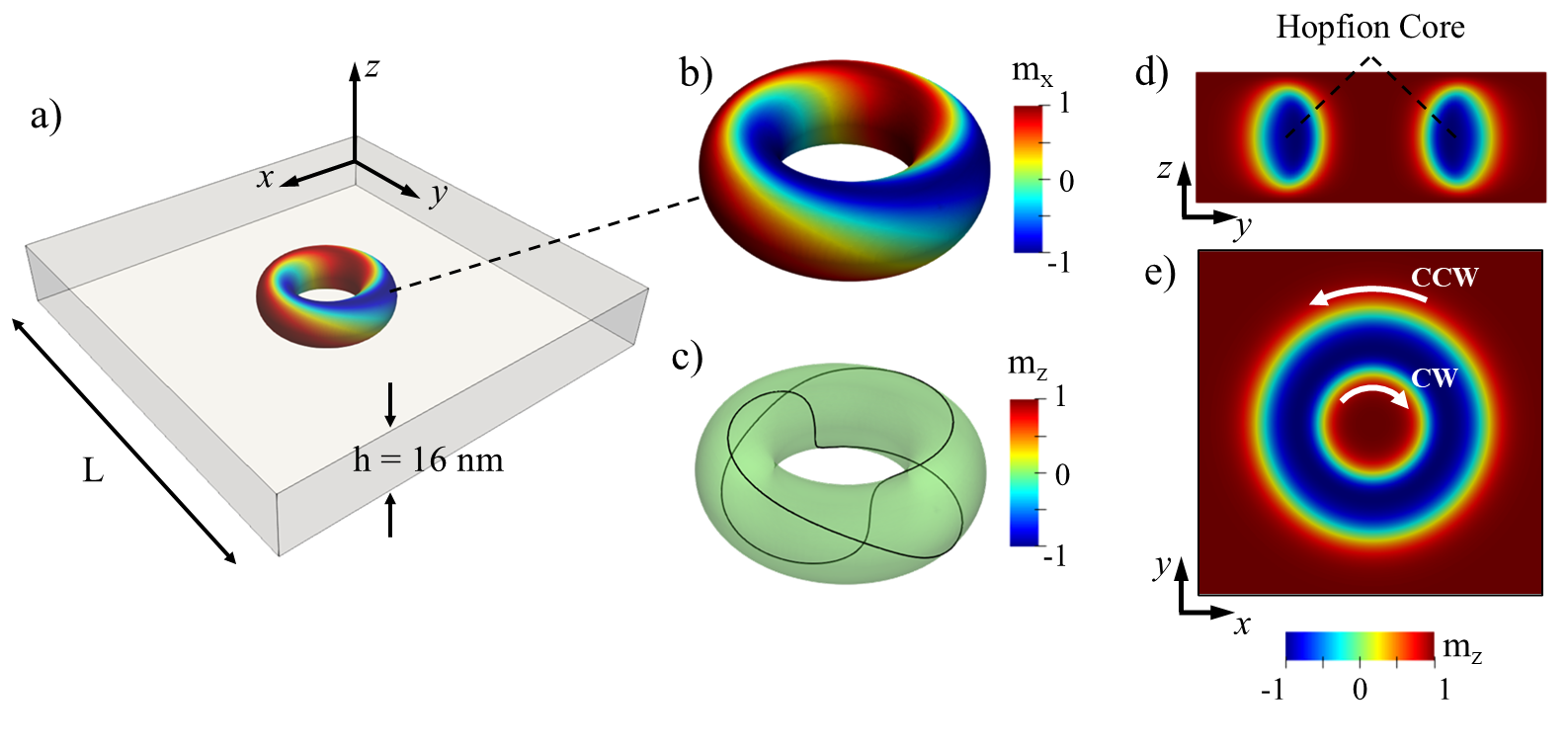}
\caption{
\textbf{Three-dimensional structure of a magnetic hopfion in a confined ferromagnetic nanodot.} 
(a) Schematic of the simulated system with the hopfion hosted inside the nanodot (the detailed hopfion structure is shown in panel (b)). 
(b) \(m_x\) component of the hopfion magnetisation mapped on the isosurface defined by \(m_z = 0\). 
(c) Same \(m_z = 0\) isosurface with overlaid black contours denoting the region where \(m_x = 0\), highlighting the internal twisting of the magnetisation. 
(d) Cross-sectional view of the \(m_z\) component in the central \(yz\)-plane, showing the hopfion core polarised along the \(-z\) direction. 
(e) Cross-sectional view of the \(m_z\) component in the central \(xy\)-plane, showing that the inner domain wall rotates clockwise (CW) while the outer domain wall rotates counterclockwise (CCW).
}
    \label{fig:Figura01}
\end{figure*}

To investigate the static and dynamic properties of magnetic hopfions under external magnetic fields, we consider a three-dimensional chiral ferromagnetic nanodot. The simulated nanodot has a fixed thickness h = 16 nm and lateral dimensions L = 200, 250 and 300 nm, chosen to fully contain the hopfion structure while minimising boundary-induced distortions. The overall three-dimensional structure of the confined hopfion is illustrated in Fig.~\ref{fig:Figura01} a-c. This figure highlights the internal topology of the soliton by displaying the magnetisation components on isosurfaces and cross-sections of the nanodot. In particular, the isosurfaces of \(m_z=0\) reveal the twisted linkage of the magnetisation field lines, while the cross-sectional views emphasise the polarised hopfion core and the opposite chirality of the inner and outer domain walls (Figs.~\ref{fig:Figura01}d and ~\ref{fig:Figura01}e, respectively).
Simulations were performed using the GPU-accelerated software \textsc{MuMax3}~\cite{vansteenkiste2014design}, which integrates the Landau–Lifshitz–Gilbert (LLG) equation within the micromagnetic approximation. To comprehensively capture both the equilibrium and dynamic behavior of magnetic hopfions, we employed two complementary computational strategies. Static configurations were obtained via energy minimization using the conjugate gradient method, enabling us to characterise the equilibrium response of the hopfion under out-of-plane magnetic fields. Subsequently, direct integration of the LLG equation captures the time evolution and extract eigenmodes following pulsed field excitation, including the emergence of breathing modes. All simulations were carried out at zero temperature, neglecting thermal fluctuations, to isolate the intrinsic deterministic dynamics of the system.

The material parameters were chosen to emulate prototypical B20-type chiral magnets, such as MnSi, in accordance with previous theoretical and numerical studies~\cite{wang2019current}. Specifically, we used an exchange stiffness \( A_{\mathrm{ex}} = 0.16\,\mathrm{pJ/m} \), saturation magnetisation \( M_s=1.51 \times 10^5\,\mathrm{A/m} \), bulk uniaxial anisotropy \( K_b = 41\,\mathrm{kJ/m^3} \), and bulk Dzyaloshinskii–Moriya interaction (DMI) strength \( D_b=0.115\,\mathrm{mJ/m^2} \). To \Rub{mimic} strong perpendicular interfacial anisotropy, the magnetisation at the top and bottom layers of the simulated structure was fixed along the \( +z \)-direction, following established methodologies~\cite{bo2021spin,tai2018static,li2022mutual}. This boundary condition effectively reproduces the presence of high-anisotropy surface layers, which are commonly required for the stabilisation of confined hopfions~\cite{khodzhaev2022hopfion,raftrey2021field,wang2019current, liu2018binding}. Finally, the nanodot was discretised with a uniform mesh of \(0.5 \times 0.5 \times 0.5\,\mathrm{nm}^3\), smaller than the exchange length and DW width, ensuring accurate resolution of the hopfion texture.
To initialise hopfion-like textures, we adopted  stereographic ansatz inspired by earlier theoretical constructions of hopfions~\cite{HIETARINTA199960}. This ansatz—explicitly defined in equations~(\ref{eq:ansatz_mx})–(\ref{eq:ansatz_mz})—has proven to be a reliable approximation of the hopfion structure and has been successfully used in previous stabilization studies~\cite{wang2019current}.

\begin{equation}
m_x = \frac{4 r' \left[ 2 z' \sin \phi + \cos \phi \left( r'^2 + z'^2 + 1 \right) \right]}{\left( 1 + r'^2 + z'^2 \right)^2},
\label{eq:ansatz_mx}
\end{equation}

\begin{equation}
m_y = \frac{4 r' \left[ -2 z' \cos \phi + \sin \phi \left( r'^2 + z'^2 + 1 \right) \right]}{\left( 1 + r'^2 + z'^2 \right)^2},
\label{eq:ansatz_my}
\end{equation}

\begin{equation}
m_z = 1 - \frac{8 r'^2}{\left( 1 + r'^2 + z'^2 \right)^2},
\label{eq:ansatz_mz}
\end{equation}

where \( r' = (e^{R/w_R} - 1)/(e^{r/w_R} - 1) \) and \( z' = \left( z / |z| \right) \left( e^{|z|/w_h} - 1 \right)/(e^{h/w_h} - 1) \). Parameters such as \(R\), \(w_R\), \(h\), and \(w_h\) define the geometric dimensions of the hopfion profile. In these equations the radius \(R\) is defined such that \(m_z(r = R, z = 0) = -1\), while the height \(h\) marks the maximum extent of the hopfion's vertical spread, determined where \(m_z(r = R, z = h) = 1/9\). The parameters \(w_R\) and \(w_h\) denote the thickness of the hopfion walls in the radial and vertical directions, respectively, marking the gradient of the magnetization transition from \(m_z = +1\) to \(m_z = -1\). These parameters set the initial geometry, which was subsequently relaxed to the nearest stable configuration under the chosen boundary conditions and applied fields.
\section{Results}
\subsection{Magnetic field tuning of hopfion size}
\begin{figure}[]
\includegraphics[width=1\linewidth]{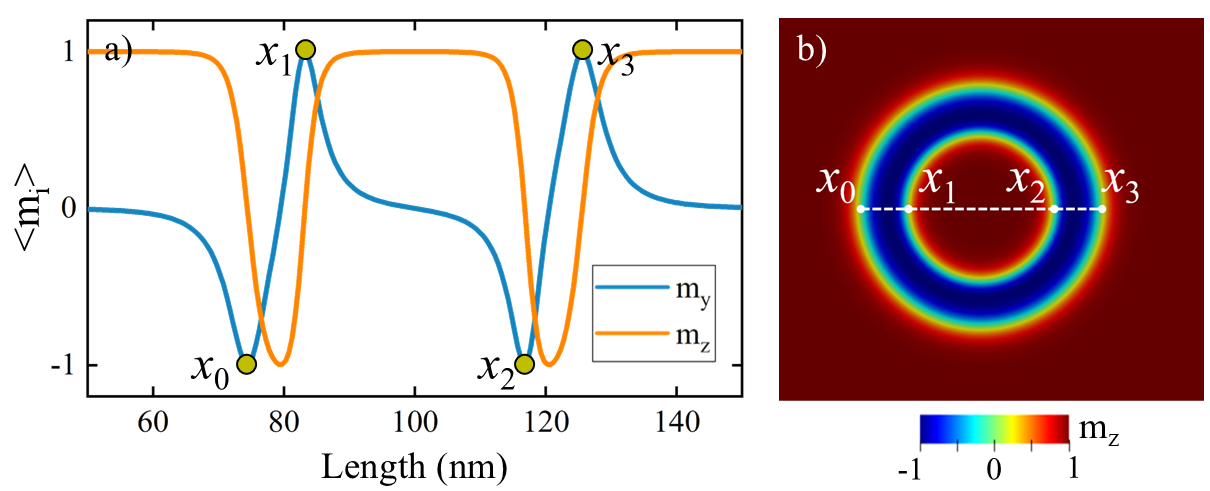}
\caption{
\textbf{Spatial profile of the hopfion texture along the central plane of the system.}
(a) Line profiles of \( m_y \) (blue) and \( m_z \) (orange) extracted from the mid-plane (\( z = 8\,\mathrm{nm} \)), revealing four distinct domain walls located at positions \( x_0 \), \( x_1 \), \( x_2 \), and \( x_3 \). The in-plane component \( m_y \) exhibits a complete \( 2\pi \) rotation, consistent with the topological structure of a hopfion. 
(b) Two-dimensional map of \( m_z \) in the same plane, with markers indicating the positions of \( x_1 \), \( x_2 \), and \( x_3 \), corresponding to the domain wall centers identified in (a). 
This spatial mapping establishes a correspondence between the one-dimensional magnetisation profile and the full two-dimensional magnetic texture. The positions \( x_0 \), \( x_1 \), \( x_2 \), and \( x_3 \) are used to define the hopfion diameter \( D_H = x_2 - x_1 \) and the internal width \( w_H = x_3 - x_2 = x_1 - x_0 \).}
    \label{fig:Figura02}
\end{figure}

We first examine the response of the hopfion structure to externally applied out-of-plane magnetic fields, $H_{\text{ext}}$. For each lateral nanodot size \( L \), equilibrium configurations were obtained by minimising the total magnetic energy via the conjugate gradient method, using as an initial state the stereographic analytical ansatz defined in equations~(\ref{eq:ansatz_mx})~-~(\ref{eq:ansatz_mz}).\\
To quantify the spatial deformation of the hopfion, we introduce two characteristic length scales: the hopfion diameter \( D_H \) and its internal width \( w_H \). As illustrated in Fig.~\ref{fig:Figura02}, the diameter \( D_H = x_2-x_1 \) is defined as the distance between the outer domain wall positions \( x_1 \) and \( x_2 \), while the internal width \( w_H = x_3 - x_2 = x_1 -x_0 \) , which measures the separation between inner and outer domain walls. These were extracted from magnetisation profiles along a horizontal cross-section at the mid-plane (z=8 nm) of the nanodot.

\begin{figure}[H]
    \centering
    \includegraphics[width=1\linewidth]{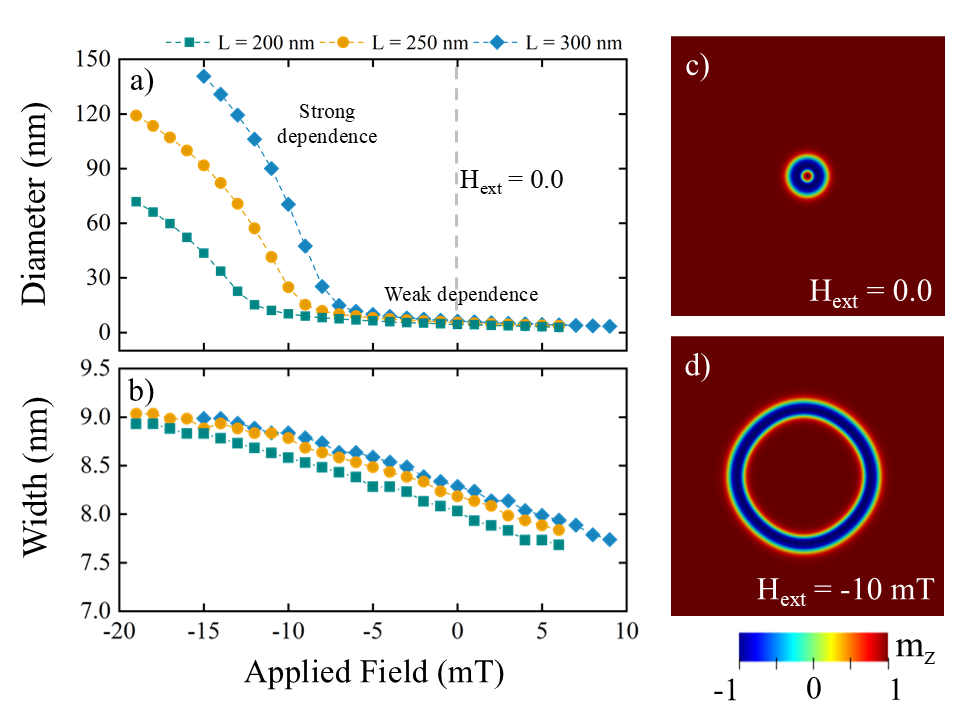}
\caption{
\textbf{Magnetic field dependence of hopfion diameter and width.}
(a) Dependence of the hopfion diameter on the externally applied static magnetic field \(B_z\), evaluated for three system sizes with lateral dimensions \(L = 200\,\mathrm{nm}\), \(250\,\mathrm{nm}\), and \(300\,\mathrm{nm}\).
(b) Corresponding dependence of the hopfion width, which remains approximately constant and weakly dependent on both field and system size.
(c,d) Cross-sectional maps of \(m_z\) in the mid-plane (\(z = 8\,\mathrm{nm}\)) for the largest lateral size (\(L = 300\,\mathrm{nm}\)), comparing the zero-field case (c) with the configuration under a negative field \(B_z = -10\,\mathrm{mT}\) (d). The hopfion expands laterally under negative field without significantly altering its width.} 
    \label{fig:Figura03}
\end{figure}

We next investigate how the hopfion responds to a static magnetic field \( H_{\mathrm{ext}} \) applied along the \( z \)-axis. The field is varied from \(-20\,\mathrm{mT}\) to \(+10\,\mathrm{mT}\) in increments of \(1\,\mathrm{mT}\). The internal structure of the hopfion (hopfion core) is polarized along the \( -z \)-direction; thus, positive fields are applied antiparallel to the core magnetization, while negative fields are parallel to it. Figure~\ref{fig:Figura03}a shows the resulting variation of the hopfion diameter \( D_H \) as a function of \( H_{\mathrm{ext}} \) for each nanodot size. Two distinct regimes can be identified. Under weak fields, \( w_H \) exhibits a weak dependence on \( H_{\mathrm{ext}} \), remaining nearly constant and reflecting a high degree of structural stability, see Fig.~\ref{fig:Figura03}b. For stronger negative fields, however, \( D_H \) shows a strong dependence on the applied field and increases sharply, revealing an outward expansion of the outer shell of the hopfion. In contrast, under positive fields, \( D_H \) decreases slightly and then saturates, suggesting that further contraction is limited by internal energetic constraints. This behaviour can be attributed to the topological protection associated with the Hopf index, $H$: a substantial reduction of the internal core would entail a topological phase transition. 

The internal width \( w_H \) remains nearly unchanged across the entire range of applied fields (Fig.~\ref{fig:Figura03}b). In all cases, its variation does not exceed \( 1.5\,\mathrm{nm} \), highlighting the rigidity of the internal domain wall configuration. This clear decoupling between the hopfion diameter and internal width indicates that the applied field predominantly perturbs the outer shell, leaving the core structure largely unaffected. Such an asymmetric and non-linear response reflects both the underlying topology of the hopfion and the competing energetic contributions that stabilise it. While the outer domain walls shift their position to minimise the Zeeman energy, the inner domain walls—topologically linked and energetically anchored by exchange and anisotropy—resist deformation. This observation points towards an equilibrium configuration governed by the competition between internal repulsive interactions and the external magnetic pressure. Additional insight is provided by the spatial maps shown in Figs.~\ref{fig:Figura03}c and \ref{fig:Figura03}d, which depict the out-of-plane magnetisation \( m_z \) at the mid-plane for two representative field values: \( H_{\text{ext}} = 0 \) and \( H_{\text{ext}} = -10\,\mathrm{mT} \). These maps reveal a lateral expansion of the hopfion under negative field, with its internal structure remaining intact. The well-defined boundary corresponding to the \( m_z = 0 \) isocontour supports the definition of the hopfion diameter and confirms the robustness of our geometric descriptors. Importantly, for the smaller lateral sizes, \(L = 200\,\mathrm{nm}\) and \(250\,\mathrm{nm}\), a magnetic field larger than \(H_\text{ext} \geq 6\,\mathrm{mT}\) induces a topological phase transition, driving the system into the uniformly polarised state along the \(+z\) direction.
\subsection*{Analytical model of domain wall interactions}
Micromagnetic simulations show that, while the hopfion diameter D$_H$ varies significantly with the applied out-of-plane magnetic field, the internal core width \( w_H \) remains essentially constant across the explored field range (Fig.~\ref{fig:Figura03} panels a and b). This rigidity of \( w_H \) suggests that the internal DW structure behaves as a strongly bound entity, largely insensitive to moderate field perturbations. To capture the origin of this behaviour, we construct a simplified analytical model for the interaction between the inner pair of DWs (DW$_2$ and DW$_3$ located at positions $x_2$ and $x_3$ respectively, see Fig.~\ref{fig:Figura02}) that delimit the hopfion core.
Although hopfions are inherently three-dimensional, the essential features of their lateral dynamics can be captured by analysing a central horizontal line cut across the mid-plane of the nanodot. From simulation data, we observe four distinct DWs along this cross-section, defined by zero crossings in the \( m_z \) component. These walls can be characterised by their winding numbers \( w = \pm \tfrac{1}{2} \), which quantify the topological rotation of the magnetisation vector in the plane. The total winding number along the 1D cross-section is given by:
\[
W=\frac{1}{2\pi}\int w(x) \,dx=\frac{1}{2\pi}\int \nabla_x \phi(x) \, dx,
\]

where \( \phi(x) \) is the azimuthal angle of the in-plane spin orientation at position \( x \). From micromagnetic simulations, we extract the winding number configuration of the four domain walls as:

\[
\vert\mathrm{DW}_0,\mathrm{DW}_1,\mathrm{DW}_2,\mathrm{DW}_3\rangle = \left\lvert \tfrac{1}{2}, \tfrac{1}{2}, \tfrac{1}{2}, \tfrac{1}{2} \right\rangle.
\]
In this configuration, the pair of DWs (DW\(_2\)-DW\(_3\) and DW\(_0\)-DW\(_1\) ) carries the same topological charge, resulting in a repulsive exchange interaction. Upon application of an external field, the region between DW$_2$ - DW$_3$ and DW$_0$ - DW$_1$ becomes energetically coupled to the field through the Zeeman term. The exchange and Zeeman energies act in cooperation: both favor increasing the separation between walls ($H_\text{ext}<0$). To quantify this, we consider a one-dimensional ansatz describing the azimuthal angle profile of two interacting domain walls as:

\begin{align}
\phi \left(x\right) = \sum_{i=1}^{2} \phi_i \left( x \right) = 2 \arctan \mathrm{exp} \left[ q_2\frac{ \left( x - x_2  \right)}{\Delta} \right]\\
+2 \arctan \mathrm{exp} \left[ q_3\frac{ \left( x - x_3 \right)}{\Delta} \right],
\tag{1}
\label{eq:1}
\end{align}
where \( q_2, q_3 \) are the winding numbers associated with the DWs located at \(x_2,x_3 \) positions respectively. $\Delta$ the DW-width and it is assumed to be the same for both DWs and defined as $\Delta=\sqrt{A/K_b}$. For identical DWs, the total exchange energy becomes:

\[
\Delta E_{\mathrm{exc}}= A \, \int^{+ \infty}_{-\infty} {\left(\sum_i \partial_x\phi_i\left(x\right) \right)}^2 dx,
\tag{2}
\label{eq:2}
\]
and the interaction energy among the two DWs can be expressed as:

\[
\Delta E^{2,3}_{\mathrm{exc}}=\frac{2q_2q_3 A}{\Delta^2} \int^{+\infty}_{-\infty}\text{sech}\left(\frac{x-x_2}{\Delta}\right)\text{sech}\left(\frac{x-x_3}{\Delta}\right)dx.
\tag{3}
\label{eq:3}
\]

\begin{figure}[!ht]
\includegraphics[width=8cm]{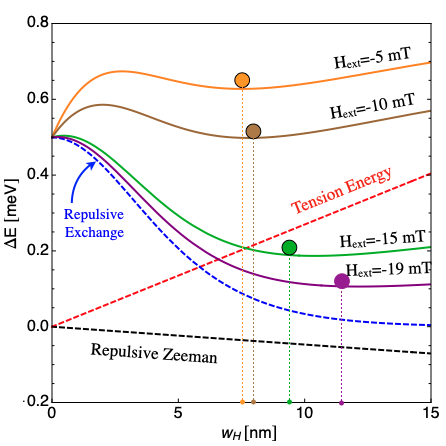}
\caption{
\textbf{Effective energy model for the variation of the width of hopfion under external magnetic field.}
Total energy curves $\Delta E(w_H)$, composed of exchange, Zeeman and DW tension contributions, as a function of domain walls (DWs) separation \(w_H\), for several values of \(H_{\mathrm{ext}}\). 
Dashed Blue curve corresponds to the dependence of the exchange repulsive interaction among DW$_{2(0)}$ and DW$_{3(1)}$. Red Dashed line shows the increase of the tension energy as a function of the expansion of \( w_H \) and black dashed lines represents how the Zeeman energy decreases as the separation among the DWs increase (\( w_H \)). The filled dots in the curves show local minima indicating stable width of the hopfion configurations. From analytical calculations, \( w_H \) for the investigated fields in micromagnetic simulations is bounded between $11$ nm for $-19$ mT to $7.2$ nm for $-5$ mT showing good quantitative agreement with micromagnetic simulations.}
\label{fig:Figura04}
\end{figure}

Note that this expression includes contributions from both: self-energy and mutual interaction. The total exchange energy, which separates these contributions, reads:

\[
\Delta E^{2,3}_{\mathrm{exc}}=\frac{4 A}{\Delta}+\frac{4 q_2q_3 A\left(x_2-x_3\right)}{\Delta^2} \text{csch} \left(\frac{x_2-x_3}{\Delta} \right),
\tag{4}
\label{eq:4}
\]

with the total topological charge of the pair $q = q_2 + q_3 = 1$. This implies a net repulsion between the DW$_2$ - DW$_3$ due to exchange coupling. Simultaneously, the external magnetic field $H_{\mathrm{ext}}$ introduces a Zeeman energy that acts promoting the separation between them.
Furthermore, a DW behaves as an elastic interface among two magnetic domains with opposite polarization with surface energy density $\sigma$ as a result of the balance among the exchange and anisotropy $\sigma=4\sqrt{A K_b}$ in the absence of DMI (small compared to the exchange stiffness in our case of study). As illustrated in Fig. ~\ref{fig:Figura01}, each DW forms a closed ring of circumference 2$\pi x_i$ so, its tension per unit of thickness is 2$\pi\sigma x_i$. This reveals that the tension cost depends only on translation of the DW-pair (DW$_{2(0)}$-DW$_{3(1)}$). For analytical tractability, we assume that DW$_2$ remains fixed while DW$_3$ is displaced outward, so that the only relevant collective coordinate describing the shell dynamics is its thickness \( w_H \). The total energy of the system is given by:
\begin{align}
\Delta E(w_{\text{H}})=\mu_0 M_{\mathrm{s}} H_{\mathrm{ext}} w_H+4\pi\sigma w_H+\frac{4A w_H}{\Delta^2} \text{csch} \left( \frac{w_H}{\Delta} \right),
\tag{5}
\label{eq:5}
\end{align}

where \( \mu_0 \) is the vacuum permeability and \( M_{\mathrm{s}} \) is the saturation magnetisation. The competition between repulsive exchange and Zeeman and compressive DW-tension term determines the equilibrium separation between the DW$_2$ and DW$_3$. Note that this can be directly extrapolated to DW$_0$ and DW$_1$. For each value of the applied field, this equilibrium corresponds to a minimum in \( \Delta E(w_{\text{H}}) \), which can be found numerically.
As illustrated in Fig.~\ref{fig:Figura04}, the total energy profile \(\Delta E(w_H)\) exhibits a pronounced minimum for each value of the applied magnetic field, indicating the presence of a stable equilibrium separation between the DWs and thus defined a stable hopfion width, $w_H$. This minimum shifts as a function of \(H_{\text{ext}}\), reflecting a competition between repulsive exchange and Zeeman interaction and the contracting DW-tension effect. For negative fields, aligned parallel to the internal magnetisation within the hopfion shell, the energy minimum shifts towards larger separations, favoring a slight expansion of the hopfion width. However, for positive external magnetic fields, the Zeeman contribution enhances the compression of the hopfion width, whereas the exchange interaction prevents further contraction because of the repulsion among the DWs conforming the hopfion shell. This field-dependent behaviour explains the robust rigidity of the shell of the hopfion shown in Fig.~\ref{fig:Figura03}b. For magnetic fields $H_{\text{ext}}=-19$ mT to $H_{\text{ext}}=-5$ mT the width size is bounded between $7.2$ nm and $11$ nm for the respective applied magnetic fields. Magnetic fields reaching positive values, it can be seen from Fig.~\ref{fig:Figura04} that the location of the minimum does not suffer substantial change and it gradually and smoothly shifts towards smaller \( w_H \) values consistent with the numerical simulations, see Fig.~\ref{fig:Figura03} panel b). These results, obtained through the analytical model, confirm that this approach quantitatively reproduces the same trend observed in the micromagnetic simulations, thereby supporting the validity of the theoretical formulation. 

\begin{figure}[!ht]
    \centering
    \includegraphics[width=0.9\linewidth]{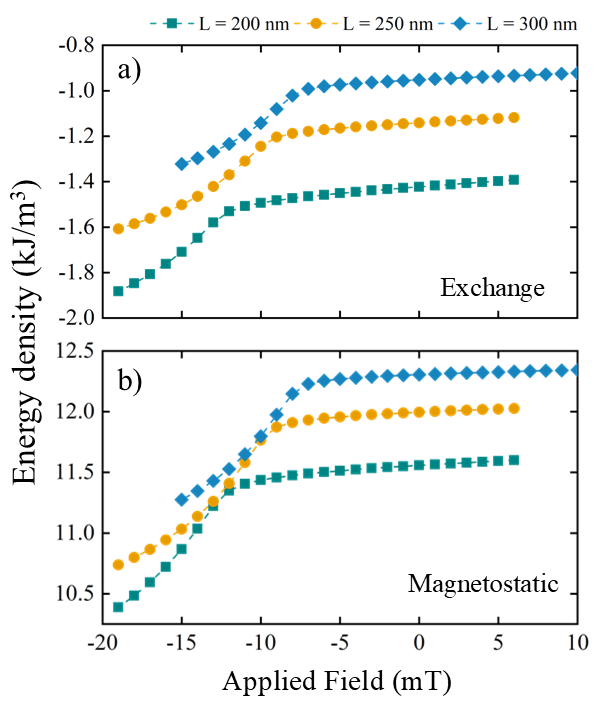}
    \caption{Exchange (\(E_{\mathrm{exch}}/V\)) and magnetostatic (\(E_{\mathrm{mag}}/V\)) energy densities of the nanodot (lateral size \(L = 200,\,250,\) and \(300\,\mathrm{nm}\)) as a function of the applied out-of-plane magnetic field. The values are obtained from micromagnetic simulations and refer to the total system that hosts the hopfion. The energies are normalised by the corresponding nanodot volume to enable like-for-like comparison across different \(L\). Negative fields promote a lateral expansion of the hopfion, which reduces spatial gradients of \(\mathbf{m}\) and effective magnetic charges, leading to a correlated decrease in both exchange and magnetostatic energy densities. The abrupt drop observed for sufficiently large negative fields reflects a threshold-driven reconfiguration of the internal domain-wall structure, consistent with the trends and interpretation discussed in Fig.~\ref{fig:Figura03}.}
    \label{fig:Figura05}
\end{figure}

Beyond reproducing the micromagnetic trends, this simplified model offers a transparent physical interpretation of the field dependent expansion and compression of the width of the hopfion, highlighting the interplay between topological DW interactions and Zeeman and DW tension effects. Furthermore, it illustrates how a fundamentally three-dimensional topological soliton can exhibit effective one-dimensional elastic behaviour, bridging complex micromagnetic dynamics with a tractable theoretical description. To further contrast the predictions of the analytical model with the full three-dimensional micromagnetic description, we computed the total exchange and magnetostatic energy densities of the simulated nanodots for different lateral sizes \(L\) under varying out-of-plane magnetic fields (Fig.~\ref{fig:Figura05}a--b). These quantities, normalised by the respective nanodot volumes, correspond to the entire system hosting the hopfion rather than to the hopfion itself, thus enabling direct comparison across different geometries. The resulting field dependence reveals that the non-linear evolution of both energy densities quantitatively encodes the same mechanism that governs the asymmetric expansion of the hopfion diameter \(D_H\) (Fig.~\ref{fig:Figura03}). As the field becomes more negative, the hopfion expands laterally: the separation between the internal domain walls ($W_1$ and $W_2$) increases, which smooths the gradients of \(\mathbf{m}\) and thereby reduces the exchange energy \(E_{\mathrm{exch}}\). At the same time, the expansion promotes three-dimensional flux closure that lowers the effective magnetic charges, leading to a decrease in the magnetostatic energy \(E_{\mathrm{mag}}\).
The abrupt change arises from the competition between exchange, Zeeman, and domain-wall tension. Exchange, manifested as the mutual repulsion of the internal domain walls, consistently drives expansion, whereas the Zeeman term compresses the hopfion under positive fields and promotes expansion under negative fields. At moderate negative fields, however, the diameter remains nearly constant due to the restoring effect of domain-wall tension. Once the Zeeman gain outweighs this rigidity, the hopfion expands, accompanied by a sharp reduction in both exchange energy \(E_{\mathrm{exch}}\) and magnetostatic energy \(E_{\mathrm{mag}}\) (Fig.~\ref{fig:Figura07}).
This non-linearity is captured by the analytical modelling (Eq. \ref{eq:5}): the interaction term \(\propto \text{csch}(D_H/\Delta)\) introduces a highly non-linear response and an energetic inflection point that reproduces the rapid transition and subsequent saturation observed numerically. Moreover, the dependence on lateral size (curves for \(L = 200, 250, 300\,\mathrm{nm}\)) highlights the role of confinement: larger nanodiscs provide greater volume to redistribute \(\nabla \!\cdot\! \mathbf{M}\) and curvature, amplifying the decrease of both \(E_{\mathrm{mag}}\) and \(E_{\mathrm{exch}}\) and yielding higher plateaux consistent with a larger \(D_H\). Taken together, these energetic trends consolidate a unified interpretation: the hopfion diameter is an elastic--topological degree of freedom governed by a balance of exchange--Zeeman--magnetostatic terms which, under negative fields, enables a stable expansion regime responsible both for the asymmetry of \(D_H(H)\) and for the geometric selectivity that underpins the breathing mode.

For consistency, the datasets presented in Fig.~\ref{fig:Figura03} and Fig.~\ref{fig:Figura05} include only hopfion configurations that preserve azimuthal symmetry. For more negative fields, the hopfion undergoes a pronounced diameter increase, eventually interacting with the square boundaries of the nanodot. This edge-induced confinement breaks the azimuthal symmetry and triggers a magnetic phase transition, leading to a distorted, non-circular geometry. The effect is most pronounced for \(L = 300\,\mathrm{nm}\), where the larger available area allows the hopfion to expand more rapidly with field, causing earlier interaction with the edges compared to \(L = 250\,\mathrm{nm}\) and \(L = 200\,\mathrm{nm}\).
\subsection*{Breathing mode dynamics}
\begin{figure}[H]
    \centering
    \includegraphics[width=1\linewidth]{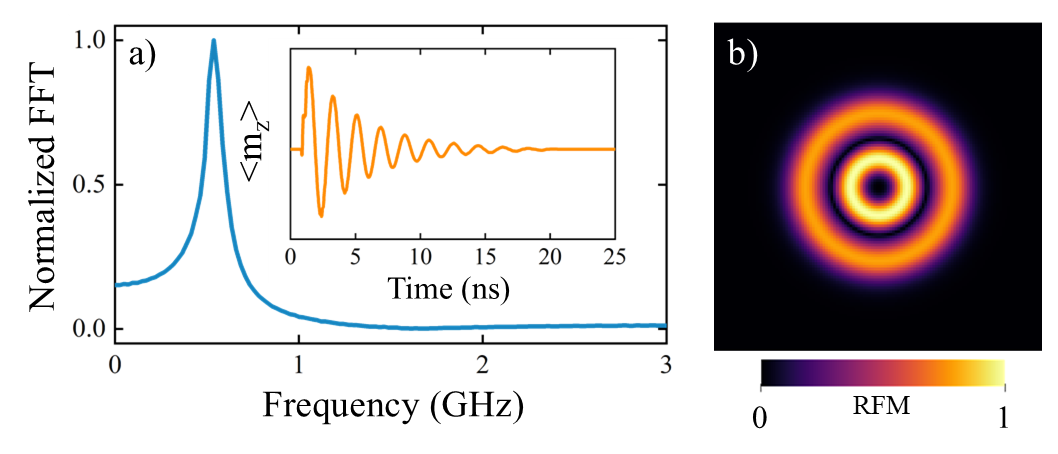}
\caption{
\textbf{Breathing mode excitation of a magnetic hopfion under combined static and dynamic magnetic fields.}
(a) FFT spectrum of the spatially averaged out-of-plane magnetisation component, \( \langle m_z(t) \rangle \), following the application of a sinc-shaped magnetic pulse along the \( z \)-axis. The system was initially stabilised under a static magnetic field of \( B_z = -9\,\mathrm{mT} \), which remained constant during the excitation. A pronounced resonant peak appears near \( 0.54\,\mathrm{GHz} \), corresponding to the breathing mode of the hopfion. The inset shows the temporal evolution of \( \langle m_z(t) \rangle \), displaying damped oscillations driven by the excitation.
(b) Spatial map of the normalised FFT amplitude in the mid-plane (\( z = 8\,\mathrm{nm} \)), exhibiting radial symmetry that confirms the breathing character of the mode.}
    \label{fig:Figura06}
\end{figure}

\begin{figure}[!ht]
    \includegraphics[width=1\linewidth]{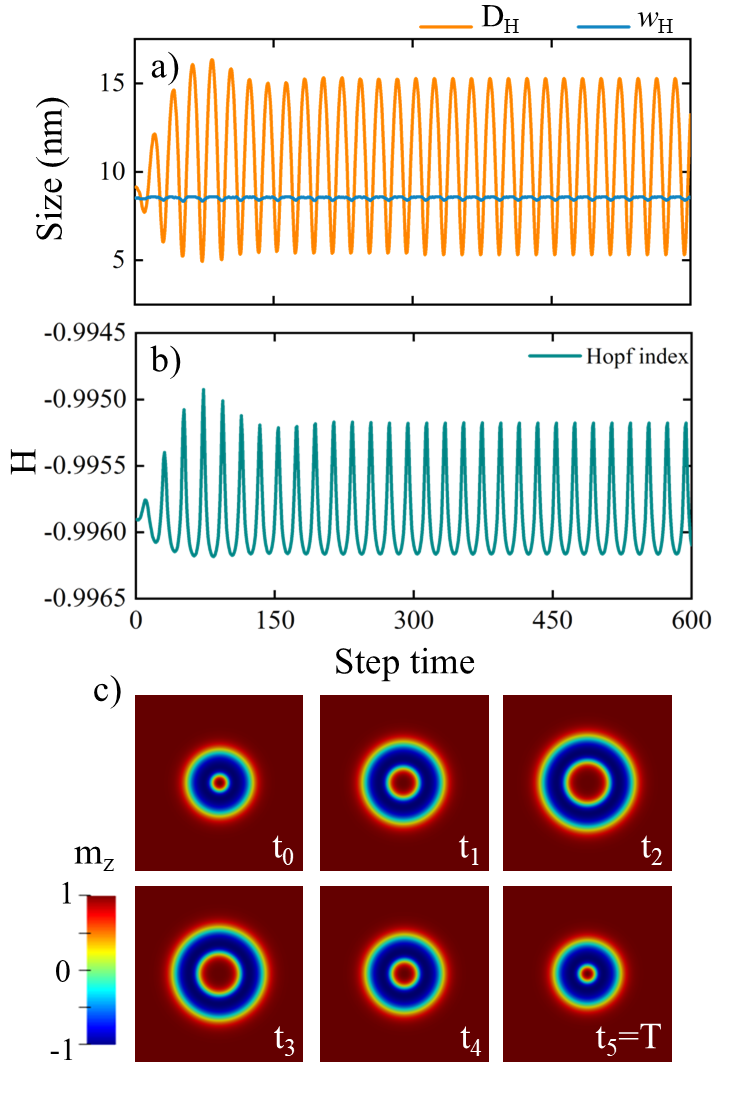}
\caption{
\textbf{Time-resolved dynamics of the hopfion breathing mode under continuous resonant excitation.} 
(a) Temporal evolution of the hopfion diameter (\(D_H\), orange) and internal width (\(w_H\), blue) over 30 oscillation cycles, showing large-amplitude periodic modulation of \(D_H\) while \(w_H\) remains nearly constant. 
(b) Corresponding time trace of the Hopf index \(H\), demonstrating topological invariance throughout the excitation within numerical precision. 
(c) Snapshots of the \(m_z\) component in the \(xy\)-plane at selected times \(t_0\)–\(t_5 = T\) (one oscillation period \(T\)), illustrating the symmetric radial expansion and contraction of the hopfion outer shell. The colour scale represents \(m_z\), from \(-1\) (blue) to \(+1\) (red). All panels correspond to a nanodot with lateral size \(L = 200\,\mathrm{nm}\).}
    \label{fig:Figura07}
\end{figure}

The observed rigidity of the internal width \( w_H \) under varying static magnetic fields, as described in the previous sections, naturally raises the question of whether this robust core structure can support localized internal excitations. In particular, the strong field-dependence of the hopfion diameter $D_H$ in contrast to the smooth variation of \( w_H \) suggests the possibility of selectively exciting dynamic deformations where the core remains static while the outer layers undergo oscillations. Motivated by this, we investigate the dynamic response of the hopfion when subject to time-dependent magnetic fields, aiming to identify and characterise potential resonant modes. To probe the natural dynamic modes of the hopfion, we apply a short magnetic pulse along the out-of-plane direction ($z$-axis), acting as a broadband excitation. This perturbation momentarily disturbs the magnetization field and triggers the intrinsic eigenmodes of the system. We then record the time evolution of the spatially averaged magnetization component $\langle m_z(t) \rangle$ over a period of several nanoseconds. Figure~\ref{fig:Figura06}a shows the frequency spectrum obtained by applying a fast Fourier transform (FFT) to the temporal signal.
The spectrum reveals a sharp and isolated peak centered at approximately $0.54\,\mathrm{GHz}$, indicating the presence of a well-defined, low-frequency excitation mode. The isolation and intensity of the peak suggest that the hopfion hosts a single dominant dynamic mode that can be excited efficiently under proper driving conditions. To further elucidate the nature of this mode, we analyze its spatial profile by computing the local oscillation amplitude of $m_z$ at each point in the system during the transient response. Figure~\ref{fig:Figura06}b displays the resulting spatial map in the mid-plane ($z = 8\,\mathrm{nm}$), where the hopfion structure is most prominent. The dynamic response is strongly localized in the radial direction, with maximal amplitude concentrated around the boundary of the hopfion core, near the domain walls where $m_z = 0$. In contrast, the central region and the exterior of the nanodot exhibit minimal dynamic activity. This spatial pattern is characteristic of a breathing mode—an oscillation where the solitonic structure expands and contracts symmetrically around its center without shifting its core position or modifying its topology.
Figure~\ref{fig:Figura07}a shows that the breathing mode is primarily manifested as an oscillation of the hopfion diameter. However, micromagnetic simulations show that the shell width \( w_H \) also oscillates at the same frequency. In the collective–coordinate picture, the hopfion's breathing mode corresponds to small oscillations of its shell width \( w_H \) around some equilibrium value $w_0$. The key quantity controlling these oscillations is the curvature, $K_w$ of the effective energy landscape obtained and presented in Eq.~\ref{eq:5} and defined as:

\begin{equation}
K_w= \left.\frac{\partial^2\Delta E}{\partial w_H^2}\right|_{w_0}\\
 =\frac{4A}{\Delta^3} \cdot \text{csch}(x) \left[
x\left( \text{coth}^2(x) + \text{csch}^2(x) \right)-2\, \text{coth}(x)
\right],
\end{equation}

where $K_w$ physically represents a restoring spring-constant which tries to avoid deviations from the equilibrium width, $w_H$. The frequency of the breathing mode is naturally defined as $w_0=\sqrt{K_w/M_w}$, where $M_w$ is the inertia associated with the coupled DWs that constitute the hopfion's shell (DW$_{2(0)}$ and DW$_{3(1)}$) and is given by: $M_w=2 (M_s t \gamma)^2\,l/ \Delta\, K_ {\perp}$ where $l=\pi D_h$ and $K_\perp$ can be estimated as $K_\perp=1/2\mu_{0}M_s^2$.
This reveals that the core and shell degrees of freedom are dynamically coupled, consistent with the analytical model where the curvature of the interaction energy defines a single collective breathing coordinate. Therefore, the analytical model allows us to directly estimate the breathing frequency, $f_0$ directly from the material parameters used in the micromagnetic simulations. Substituting the parameters used in micromagnetic simulations, we find: 

\begin{equation}
f_0=\frac{1}{2}\sqrt{\frac{K_w}{M_w}}\approx 0.56 \,\,\text{GHz},
\end{equation}

in very good agreement with the spectral peak observed in the numerical simulations ($\approx 0.54$ GHz). This confirms that the breathing mode can be quantitatively understood as a harmonic oscillation of the shell thickness governed by the curvature of the energy landscape. To elucidate the resonant nature of the mode and quantify its effect on the hopfion geometry, we perform a second set of simulations in which we apply a sinusoidal magnetic field along the $z$-axis with frequency matched to the previously observed peak (0.54 GHz). The amplitude of the field is set to a small value ($0.5\,\mathrm{mT}$) to remain within the linear regime. We then monitor the temporal evolution of both the hopfion diameter $D_H(t)$ and internal width $w_H(t)$ over 120 oscillation cycles. Figure~\ref{fig:Figura07} a) shows the time traces of these two quantities during the first 30 cycles. The hopfion's diameter exhibits a clear sinusoidal oscillation at the driving frequency, with an amplitude of several nanometres, while the internal width \(w_H = x_1 - x_0\) remains nearly constant throughout the entire simulation. This demonstrates that the excitation selectively drives the outer shell of the hopfion, with \(w_H\) exhibiting only minor fluctuations around its equilibrium value. Moreover, the phase coherence and stability of the oscillation over many cycles indicate that the mode is spectrally well defined and only weakly damped. From the frequency spectrum shown in Fig.~\ref{fig:Figura06}, we estimate a quality factor of $Q \approx 3.96$.

Importantly, during the breathing dynamics, the overall topology of the hopfion remains preserved. The Hopf index, $H$, which measures the topological entanglement of the magnetization field lines, remains invariant to within numerical precision throughout the simulation (see Fig. \ref{fig:Figura05} b). This is consistent with the absence of singularities or discontinuities in the field, ensuring that the breathing mode represents a continuous and topologically trivial deformation of the texture. To our knowledge, this is the first direct report of a breathing mode in magnetic hopfions, and it highlights the existence of internal low-frequency excitations in three-dimensional solitonic textures. Unlike skyrmions\cite{kim2014breathing}, whose breathing modes often suffer from mode mixing or edge-induced distortion, the three-dimensional topology of the hopfion enables a spectrally isolated and radially symmetric oscillation that preserves topological integrity over many cycles. This discovery is not only of fundamental interest in the study of magnetization dynamics and topology, but also holds promise for potential applications. For example, resonant control of such modes could enable selective manipulation of hopfion states using AC-magnetic fields or spin-torque oscillators. Additionally, the breathing frequency may serve as a spectral fingerprint for detecting the presence and size of hopfions in experimental systems using ferromagnetic resonance (FMR) techniques.
\section*{Discussion}
The discovery of a resonant breathing mode in magnetic hopfions reveals a previously unknown dynamic degree of freedom in three-dimensional topological solitons, offering opportunities for both fundamental studies and spintronic applications. Our results show that the hopfion diameter can be reversibly tuned by an out-of-plane magnetic field, while the internal core width remains essentially unchanged. This geometric decoupling, absent in two--dimensional skyrmions, enables coherent radial oscillations of the outer shell without perturbing the topological core, establishing hopfions as solitons with an unprecedented separation between structural and topological responses.
Furthermore, by combining high-resolution micromagnetic simulations with an analytical framework based on DW interactions and self-DW tension energies, we demonstrate that the field dependence of the hopfion width arises from the competition between exchange-mediated and Zeeman repulsion and DW-tension compression. This balance defines a field-dependent equilibrium separation of the core of the hopfion. The model not only reproduces the trends observed in simulations but also provides a physically transparent picture of how a fully three-dimensional texture can exhibit effectively one-dimensional elastic behaviour. 
Dynamically, we identify a robust breathing mode that appears as a sharp sub-GHz resonance ($\approx 0.54\,\mathrm{GHz}$) in the frequency spectrum following broadband excitation. Notably, our analytical DW interaction model predicts a breathing frequency of $f_0 \approx 0.56$ GHz, in excellent agreement with the simulated resonance at $0.54$ GHz, confirming that the curvature of the interaction potential fully accounts for the observed mode. Spatial maps reveal that this mode is localized at the soliton boundary, where the oscillation amplitude is maximal, while the core barely moves. Under resonant sinusoidal driving, the hopfion diameter oscillates with nanometre-scale amplitude over hundreds of cycles without measurable degradation of the Hopf index, confirming both the coherence and the topological integrity of the mode. This degree of selectivity in exciting only the inner structure of the soliton, while leaving its core unaltered, has no precedent in lower-dimensional magnetic textures.
An additional point of note is that the hopfion breathing mode exhibits a moderate quality factor ($Q \approx 3.96$). Although this value is lower than that which could be obtained in many other magnetic solitons, the mode remains clearly identifiable and spectrally isolated. Moreover, a reduced $Q$ can also be advantageous: the broader linewidth facilitates excitation over a wider frequency range, enhances the coupling efficiency to external fields, and prevents excessive energy accumulation that could otherwise drive nonlinear distortions or instabilities.
These findings establish hopfions as dynamically rich objects whose internal modes can be accessed and manipulated without compromising stability. In practical terms, the breathing frequency could serve as a spectral fingerprint for experimental detection via ferromagnetic resonance or Brillouin light scattering, while resonant driving might enable frequency-encoded control schemes in reconfigurable spintronic architectures. The ability to modulate hopfion geometry in real time, coupled with their inherent topological protection, points to promising routes for encoding information in both spatial and spectral domains. Looking ahead, the next challenges include exploring how other internal modes---such as twisting or gyrotropic oscillations---interact with the breathing motion, and determining how finite temperature, damping, and spin currents influence mode stability. Crucially, translating these predictions into experimental realizations will require material platforms where hopfions can be generated, stabilized, and probed under controlled magnetic fields. By enabling frequency-selective control of three-dimensional topological solitons, our results provide a blueprint for the next generation of functional, energy-efficient spintronic devices.
\section*{Methods}
\subsection*{Breathing Mode Excitation}
To excite the internal breathing mode of the hopfion, we apply a short out-of-plane magnetic field pulse of the form
\[
B_z(t) = B_0 + b_0 \, \mathrm{sinc}\left(2\pi f_c (t - t_0)\right),
\]
where \( B_0 = -9\,\mathrm{mT} \), \( b_0 = 0.5\,\mathrm{mT} \), \( t_0 = 1.0\,\mathrm{ns} \), and \( f_c = 40\,\mathrm{GHz} \). A Gilbert damping constant of \( \alpha = 0.01 \) is used throughout the simulation. The time evolution of the spatially averaged \( m_z \) component is recorded with a sampling interval of \( 10\,\mathrm{ps} \), over a total duration of \( 40.96\,\mathrm{ns} \), corresponding to $2^{12}$ data. The breathing mode frequency is extracted by computing the Fast Fourier Transform (FFT) of this time series. The spatial profile of the breathing mode is obtained by computing the FFT amplitude of \( m_z(x, y, z = 8\,\mathrm{nm}, t) \) at the resonance frequency, and then plotting its magnitude.
\section*{Acknowledgements}
F.T. acknowledges support from ANID Fondecyt Iniciación 2024, project 11241113.
\section*{Author contributions statement}
F.T. conceived the study, performed the micromagnetic simulations, and analyzed the data. R.M.O. developed the theoretical model and contributed to the interpretation of results. Both authors discussed the findings and co-wrote the manuscript.
\section*{Competing interests}
The authors declare no competing interests.
\bibliography{sample}
\end{document}